\documentclass[longbibliography,twocolumn,superscriptaddress,floatfix,showpacs]{revtex4-1}

\usepackage{amsfonts, amsmath, amssymb, mathtools, physics, bbm, calc}
\usepackage[dvipsnames]{xcolor}
\usepackage{graphicx}
\usepackage{hyperref}
\usepackage{tikz}\usetikzlibrary{calc, decorations.pathreplacing}
\usepackage{ifthen}
\usepackage{cases}
\usepackage[export]{adjustbox} 

\tikzset{baseline={([yshift=-.5ex]current bounding box.center)}, line cap = round}
\tikzset{
	ncbar angle/.initial=90,
	ncbar/.style={
		to path=(\tikztostart)
		-- ($(\tikztostart)!#1!\pgfkeysvalueof{/tikz/ncbar angle}:(\tikztotarget)$)
		-- ($(\tikztotarget)!($(\tikztostart)!#1!\pgfkeysvalueof{/tikz/ncbar angle}:(\tikztotarget)$)!\pgfkeysvalueof{/tikz/ncbar angle}:(\tikztostart)$)
		-- (\tikztotarget)
	},
	ncbar/.default=0.5cm,
}
\tikzset{square left brace/.style={ncbar=0.1cm}}
\tikzset{square right brace/.style={ncbar=-0.1cm}}

\newcommand{\Sep}{\noindent\makebox[\linewidth]{\resizebox{0.5\linewidth}{1pt}{$\bullet$}}\bigskip}

\definecolor{UGentBlue}{RGB}{30,100,200}
\definecolor{UGentYellow}{RGB}{255,210,0}
\definecolor{DarkGray}{RGB}{102, 102, 102}

\hypersetup{
	colorlinks,
	linkcolor={UGentBlue},
	citecolor={UGentBlue},
	urlcolor={UGentBlue},
	pdftitle={SPT phases with frieze symmetries},
	pdfauthor={Bram Vancraeynest-De Cuiper, Jacob C. Bridgeman, Nicolas Dewolf, Jutho Haegeman, Frank Verstraete}
}

\newcommand{\mb}{\mathbb}
\newcommand{\mc}{\mathcal}
\newcommand{\msf}{\mathsf}
\newcommand{\mbm}{\mathbbm}
\newcommand{\mbf}{\mathbf}

\begin{document}

\title{One-dimensional symmetric phases protected by frieze symmetries}
\author{Bram~\surname{Vancraeynest-De Cuiper}}\email{Bram.VancraeynestDeCuiper{@}UGent.be}
\affiliation{\it Department of Physics and Astronomy, Ghent University, Krijgslaan 281, 9000 Gent, Belgium}
\author{Jacob C.~\surname{Bridgeman}}
\affiliation{\it Department of Physics and Astronomy, Ghent University, Krijgslaan 281, 9000 Gent, Belgium}
\author{Nicolas~\surname{Dewolf}}
\affiliation{\it KERMIT, Department of Data Analysis and Mathematical Modelling, Ghent University, Belgium}
\author{Jutho~\surname{Haegeman}}
\affiliation{\it Department of Physics and Astronomy, Ghent University, Krijgslaan 281, 9000 Gent, Belgium}
\author{Frank~\surname{Verstraete}}
\affiliation{\it Department of Physics and Astronomy, Ghent University, Krijgslaan 281, 9000 Gent, Belgium}

\begin{abstract}
	We make a systematic study of symmetry-protected topological gapped phases of quantum spin chains in the presence of the  frieze space groups in one dimension using matrix product states. Here, the spatial symmetries of the one-dimensional lattice are considered together with an additional `vertical reflection', which we take to be an on-site $\mathbb{Z}_2$ symmetry. We identify seventeen distinct non-trivial phases, define canonical forms, and compare the topological indices obtained from the MPS analysis with the group cohomological predictions. We furthermore construct explicit renormalization group fixed-point wave functions for symmetry-protected topological phases with global on-site symmetries, possibly combined with time-reversal and parity symmetry. En route, we demonstrate how group cohomology can be computed using the Smith normal form.
\end{abstract}

\maketitle

\section{Introduction}

Even though there is no intrinsic topological order in gapped one-dimensional quantum spin chains, the phase diagram becomes non-trivial when symmetry constraints are taken into account~\cite{Gu2009,Pollmann2010, Pollmann2012, Chen2010, Chen2011a, Chen2011b, Chen2013, Schuch2011, zeng2018quantum}. This gives rise to the well-known paradigm of \emph{symmetry-protected} topological (SPT) order. The lack of topological order in 1D can be understood as follows. Starting from the ground state of a gapped local Hamiltonian, subsequent renormalization group (RG) coarse graining steps do not alter the phase of the system~\cite{Verstraete2005}. After sufficiently many steps, the number of which is independent of system size, the state flows towards an RG fixed point exhibiting a valence bond structure~\cite{Verstraete2005, Chen2011a, zeng2018quantum}. A tensor product of unitaries on the state then turns this state in a trivial product state, ultimately proving that the state we started from is adiabatically connected to a product state with no topological order. This procedure can be made explicit by writing the state as a matrix product state (MPS)~\cite{verstraete2006, perezgarcia2007}. In this formalism one RG step is equivalent to blocking two sites and acting with an isometry on the blocked site that maximally removes local entanglement inside the block while retaining the entanglement with the rest of the system~\cite{Verstraete2005}. When symmetries are taken into account, the RG flow should not break the symmetry. The picture that arises is that the phase diagram, which in the absence of symmetries is simply connected, falls apart in distinct classes that cannot be connected by adiabatic transformations due to topological obstructions.

Chen et al.\ showed that the topological obstructions that prohibit connecting different such SPT phases originate from the fact that physical symmetries can be implemented by \emph{projective} representations of the symmetry group acting on the entanglement degrees of freedom~\cite{Chen2011a, Chen2011b, Chen2013}. This crucial insight led Chen et al.\ to a classification of SPT phases in terms of group cohomology~\cite{Brown}. More specifically, the SPT classification corresponds to the second cohomology group $H^2_{\beta}\left(\msf{G},\msf{U}_1\right)$, where $\msf{G}$ denotes the symmetry group and can contain a global on-site symmetry subgroup, time-reversal, parity or combinations thereof. $\beta$ denotes the action of $\msf{G}$ on the $\msf{U}_1$ module in case of time-reversal or parity symmetry.

A folklore example is one where the global on-site symmetry group is $\mb{Z}_2\times\mb{Z}_2$. Since the second cohomology group of $\mb{Z}_2\times\mb{Z}_2$ is $H^2\left(\mb{Z}_2\times\mb{Z}_2,\msf{U}_1\right)=\mb{Z}_2$, $\mb{Z}_2\times\mb{Z}_2$ can protect one non-trivial symmetry-protected Haldane phase~\cite{Haldane1983a, Haldane1983b, Pollmann2010, Pollmann2012}. In translationally invariant systems, the classification is refined to $H^1_\alpha\left(\msf{G},\msf{U}_1\right) \times H^2_\beta\left(\msf{G},\msf{U}_1\right)$ where $H^1$ denotes the first cohomology group~\cite{Chen2011a}.

In this paper, we demonstrate that SPT phases can also be protected by quasi-one-dimensional lattice symmetries. The symmetry groups we consider are the seven so-called \emph{frieze groups}~\cite{coxeter1961}. These are defined as being the infinite discrete subgroups of the isometries of a strip, $\text{Isom}([0,1]\times\mb{R})$. Apart from translations, the generators of the frieze groups are reflections in the horizontal or vertical direction, $\pi$-rotations (equivalent to the composition of a horizontal and vertical reflection) and glide reflections. The seven distinct frieze groups these generators give rise to are denoted by $F_0$ (only translation), $F_V$ (translation + vertical reflection), $F_H$ (translation + horizontal reflection), $F_R$ (translation + $\pi$-rotation), $F_G$ (translation + glide reflection), $F_{RG}$ (translation + $\pi$-rotation + glide reflection) and $F_{VH}$ (translation + two reflections). In case of a glide reflection, acting with this glide reflection twice is equivalent to the action of the translation generator.\par
We derive the SPT classification corresponding to these symmetries by imposing the symmetry on a general injective MPS and identifying topologically distinct ways in which this symmetry can be implemented. Here, the vertical reflection corresponds to an on-site $\mathbb{Z}_2$ symmetry of the system, which we represent explicitly as a swap of the two physical degrees of freedom associated with every local tensor. In this way, we obtain seventeen non-trivial phases. We construct explicit canonical representative MPS ans\"atze for most of these phases and give an interpretation of these phases in terms of group cohomology.\\

{\it Outline:} In Sec.~\ref{sec:MPS} we begin by providing a review of matrix product states, recapitulating the concepts of MPS injectivity, gauge transformations, the transfer matrix and the fundamental theorem of MPS. After reconsidering the implementation of symmetries in MPS and how this leads to the SPT classification of Chen et al. in Sec.~\ref{sec:SymTensors}, we present in Sec.~\ref{sec:SmithNormal} a method to compute group cohomology and explicit cocyles from the Smith normal form of the coboundary map and in Sec.~\ref{sec:ansatz} construct explicit correlation length zero MPS tensors transforming according to given cohomology classes characterizing an SPT phase. We explicitly construct our ansatz from the non-trivial 2-cocycle of $\mb{Z}_2\times\mb{Z}_2$ in Sec.~\ref{sec:Z2Z2} and find that it reduces to the fixed point cluster state dressed with a trivial dimer state. In Sec.~\ref{sec:GenAnsatz} we generalize our ansatz and argue how an MPS transforming on the physical level in some arbitrary representation of a finite symmetry group can be constructed in such a way that the virtual bond dimension is as small as possible. In Sec.~\ref{sec:frieze}, we derive the SPT classification for frieze symmetric MPS and construct canonical forms for most of these phases. We reconsider the problem of imposing time-reversal symmetry in MPS in Sec.~\ref{sec:time} and combine time-reversal with shifts over one site. Some technical details are relegated to Appendix~\ref{App:Proof}, and in Appendix~\ref{App:example} we demonstrate our algorithm to compute cocycles of $H^2_{\beta_P}(\mb{Z}_2^P,\msf{U}(1))$, the second cohomology of $\mb{Z}_2$ parity with non-trivial group action.

\subsection*{Summary of results}

In Table~\ref{Tab:Results} below we give an overview of the symmetry groups we consider and the SPT classification they give rise to.

Translation symmetry in itself does not give rise to non-trivial SPT phases. It can be shown that every translationally invariant MPS admits a uniform representation~\cite{perezgarcia2007}. The reflection in the $F_V$ symmetry group can be thought of as an on-site $\mb{Z}_2$ symmetry. We find one non-trivial phase and show that there always exists a gauge in which the MPS tensors have definite $V$-parity. The reflection in $F_H$ is equivalent to the parity considered in~\cite{Chen2011a}. Three non-trivial SPT phases are found, in accordance with~\cite{Chen2011a}, and a canonical form is found in which the tensors have definite $H$-parity, possibly at the cost of introducing non-trivial \emph{bond tensors}~\cite{jiang2015}. The same result is found for $F_R$-symmetric MPS. $F_G$ symmetry admits only the trivial phase and we show that every $F_G$-symmetric MPS can be brought in a manifestly $F_G$-invariant form. In case of the larger symmetry groups $F_{RG}$ and $F_{VH}$ there are respectively three and seven non-trivial phases.

MPS with translation and time-reversal symmetry can protect one non-trivial SPT phase, whereas if time-reversal is combined non-trivially with a shift over one site, no non-trivial phases are retained.

\begin{table}[htpb]
	\centering
	\begin{tabular}{ |c|c| }
		\hline
		Symmetry        & SPT classification    \\
		\hline
		$F_0$           & $\slash$              \\
		$F_V$           & $\mb{Z}_2$            \\
		$F_H$ (Parity)  & $\mb{Z}_2^{\times 2}$ \\
		$F_R$           & $\mb{Z}_2^{\times 2}$ \\
		$F_G$           & $\slash$              \\
		$F_{RG}$        & $\mb{Z}_2^{\times 2}$ \\
		$F_{VH}$        & $\mb{Z}_2^{\times 4}$ \\
		\hline
		$F_1^T$     & $\mb{Z}_2$            \\
		$F_2^T$ & $\slash$              \\
		\hline
	\end{tabular}
	\caption{Overview of the SPT classification of frieze group symmetries. $F_1^T$: translation invariance over one site and time reversal, $F_2^T$: translation invariance over two sites combined with time-reversal}
	\label{Tab:Results}
\end{table}

\section{Review of matrix product states}
\label{sec:MPS}
\emph{In this section we present a brief review of injective matrix product states. We focus on some key aspects that are used to derive the frieze classification below.}

\Sep

In this paper we consider bosonic spin systems. The Hilbert space is simply the tensor product of the local $d$-dimensional Hilbert spaces of the constituent spins, $\mc{H}\cong \left(\mb{C}^d\right)^{\otimes N}$. A matrix product representation of a state in $\mc{H}$ with periodic boundary conditions is of the form
\begin{equation}
	\ket{\psi} = \sum_{\{i_k\}} \Tr\left(A_1^{i_1}A_2^{i_2}\dots A_N^{i_N}\right)\ket{i_1}\ket{i_2}\dots \ket{i_N}.
\end{equation}
Such a periodic MPS can be pictorially represented as:
\begin{equation}
	\includegraphics[valign=c]{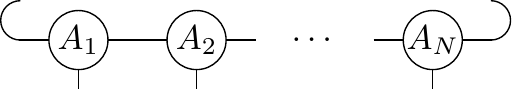}
	.
	\label{eq:GenMPS}
\end{equation}
The variational degrees of freedom are contained in the local tensors $(A_n^i)_{\alpha\beta}\in \mb{C}^D\otimes\mb{C}^D\otimes\mb{C}^d$, where $D$ is called the \emph{bond dimension}. Every state in $\mc{H}$ can be represented with a bond dimension that scales exponentially in the system size, but the power of MPS lies in the fact that ground states of gapped local Hamiltonians can be well approximated by MPS with a bond dimension that scales polynomially in the number of spins~\cite{verstraete2006}. An MPS representation of a state $\ket{\psi}$ is never unique: a \emph{gauge transformation} $A^i_n \mapsto X_n^{-1}A^i_nX_{n+1}$ clearly leaves the state invariant because the gauge tensors $X_i$ cancel on the bonds.

In case of translation invariance, it can be shown that one can always carry out a gauge transformation that brings the translationally invariant MPS in a canonical \emph{uniform} form in which $A_n\equiv A,\ \forall n$~\cite{perezgarcia2007}:
\begin{equation}
	\ket{\psi\left(A\right)}
	=
	\includegraphics[valign=c]{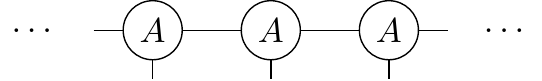}
	.
	\label{eq:UniformMPS}
\end{equation}
Gauge transformations can furthermore be used to bring the MPS parameterization in a left- or right-canonical form, characterized respectively by:
\begin{align}
	\sum_i \left(A^i\right)^\dagger A^i & = \mbm{1},
	\qquad
	\includegraphics[valign=c]{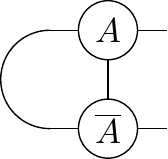}
	\quad
	=
	\quad
	\includegraphics[valign=c]{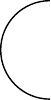}
	\ ,
	\label{eq:LeftCan}                               \\
	\sum_i A^i \left(A^i\right)^\dagger & = \mbm{1},
	\qquad
	\includegraphics[valign=c]{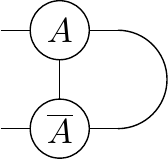}
	\quad
	=
	\quad
	\includegraphics[valign=c]{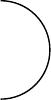}
	\ .
	\label{eq:RightCan}
\end{align}

We define the \emph{transfer matrix} $\mb{E}$ as
\begin{equation}
	\mb{E} = \sum_i A^i \otimes \bar{A}^i=
	\includegraphics[valign=c]{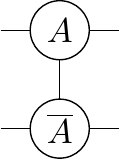}
	\label{eq:TransferM}
	.
\end{equation}
The transfer matrix captures all the relevant information about the entanglement and correlations of the state. Moreover, the transfer matrix determines the MPS uniquely up to a local change in basis. This follows from the observation that the transfer matrix defines a completely positive (CP) map where the local MPS tensors play the role of Kraus operators, combined with the fact that a Kraus decomposition of a CP map is unique up to unitary equivalence~\cite{nielsen2002quantum}.

If the matrices $\left\{A^i|i=1,...,d\right\}$ generate (via linear combinations and products) the entire $D\times D$ matrix algebra, the MPS is said to be \emph{injective}. In that case, the transfer matrix (interpreted as an $D^2\times D^2$ matrix from the left pair of indices to the right pair) has a unique eigenvalue of largest magnitude that in an appropriate normalization of the MPS tensors can be taken to be one. Moreover, when the MPS is in a left - or right canonical form, the corresponding eigenvector is $\mbm{1}$ as follows from (\ref{eq:LeftCan}-\ref{eq:RightCan}).

By far the most important property of injective MPS is that it satisfies the requirements of the \emph{fundamental theorem} of MPS: two injective uniform MPS defined by local tensors $A^i$ and $B^i$ describe the same state $\ket{\psi\left(A\right)}\sim\ket{\psi\left(B\right)}$ if and only if there is a gauge transformation $X$ and a phase $\theta$ that intertwines the two tensors: $A^i=e^{i\theta}X^{-1}B^iX$. If $A$ and $B$ are simultaneously in left (or right) canonical form, then $X$ can be chosen to be unitary. Furthermore, $e^{i\theta}$ is uniquely defined, whereas $X$ is only defined up to an overall scaling. Put differently, for $B^i = A^i$, the relation $A^i = e^{i\theta}X^{-1}A^iX$ implies $\theta=0 \mod 2\pi$ and $X=c\mbm{1}$ for some $c \in \mathbb{C}$, as follows readily from the definition of injectivity.

\section{Explicit symmetric tensors}
\label{sec:SymTensors}

\emph{In this section we review how symmetries are implemented in the tensor network language and discuss how the SPT classification arises from the projective representation of the underlying symmetry group. We present an algorithm to compute group cohomology and explicit cocycles using the Smith normal form and give an MPS ansatz that realizes the SPT phases classified by a given 1- and 2-cocycle. We work out this ansatz for the non-trivial 2-cocycle of $\mb{Z}_2\times\mb{Z}_2$ and demonstrate that this representative MPS is the tensor product of the correlation length zero cluster state and a trivial dimer state. We then discuss how an MPS can be constructed for every possible physical representation of a given finite symmetry group such that the bond dimension is minimal.}

\Sep

If an injective translationally invariant MPS $\ket{\psi\left(A\right)}$ is invariant under the action of a unitary on-site symmetry transformation, i.e. $U_g^{\otimes N}\ket{\psi\left(A\right)}\sim \ket{\psi\left(A\right)}$ for all $g \in \msf{G}$, there must exist, for every $g \in \mathsf{G}$, a phase $\varphi(g)$ and a gauge transformation $X_g$ such that
\begin{equation}
	\sum_j \left(U_g\right)_{ij}A^j = e^{i\varphi(g)} X^{-1}_g A^i X_g.
	\label{eq:TransformTensor}
\end{equation}
This equation can only admit solutions if $U_g$ forms a linear unitary representation of the on-site symmetry group $\msf{G}$. If the MPS is in either left or right canonical form, the gauge matrices $X_g$ can be chosen to be unitary. Because of the overall scale freedom in how they are determined, it follows that they only need to constitute a projective representation of $\msf{G}$, i.e. they form a representation of $\msf{G}$ up to phase:
\begin{equation}
	X_gX_h = e^{i\omega(g,h)}X_{gh}.
\end{equation}
Here, the $\omega(g,h)$'s satisfy the well-known 2-cocycle equations \begin{equation}
	\omega(g,h) + \omega(gh,k) = \omega(g,hk) + \omega(h,k) \quad \mod 2\pi,
	\label{eq:2cocycle}
\end{equation}
expressing associativity of the multiplication of the gauge matrices $X_g$. Solutions to this constraint are called 2-cocyles. The phase $\varphi(g)$, on the other hand, constitutes a one-dimensional linear representation of the symmetry group:
\begin{equation}
	\varphi(g) + \varphi(h) = \varphi(gh) \quad \mod 2\pi.
\end{equation}
The scale freedom in determining gauge transformations implies that the matrices $X_g$ can be replaced with an equivalent choice of the form $X_g\mapsto e^{ i \gamma(g)}X_g$. Under such a redefinition the cocycles transform according to
\begin{equation}
	\omega(g,h) \mapsto \omega(g,h) + \gamma(g) + \gamma(h) -\gamma(gh).
\end{equation}
Hence, in the classification of projective representations labeled by 2-cocycles, these redefinitions have to be modded out, giving rise to equivalence classes of projective representations. Cocycles $\omega$ of the form $\omega(g,h) = \gamma(g) + \gamma(h) -\gamma(gh)$, which are equivalent to the choice $\omega(g,h)=1$, are called coboundaries. The equivalence classes $[\omega]$, defined by $[\omega] = [\omega'] \iff \omega(g,h) = \omega'(g,h) + \gamma(g) + \gamma(h) -\gamma(gh)$, are exactly classified by the second cohomology group of $\msf{G}$ with respect to $\msf{U}_1$, $H^2\left(\msf{G},\msf{U}_1\right)$. In the context of group cohomology, the one-dimensional linear representation $\varphi(g)$ is referred to as a 1-cocycle and is correspondingly characterized by $H^1\left(\msf{G},\msf{U}_1\right)$. Chen et al. showed that every choice of a 1-cocycle $\varphi\in H^1\left(\msf{G},\msf{U}_1\right)$ and cohomology class $\left[\omega\right]\in H^2\left(\msf{G},\msf{U}_1\right)$ gives rise to a distinct SPT phase~\cite{Chen2011a}.

In case of a finite symmetry group, the second cohomology group is always $\mb{Z}_{d_1}\times\mb{Z}_{d_2}\times ... \times\mb{Z}_{d_N}$ for topological indices $d_1, d_2,...,d_N$, giving rise to a finite number of SPT phases. Explicit representative 2-cocycles can be obtained by writing the 2-cocycle condition as a linear system modulo $2\pi$ and solving it using the Smith normal form, as explained below in Sec.~\ref{sec:SmithNormal}.\\

In case the symmetry group contains time-reversal or parity transformations, the above picture has to be modified as follows.

Since the $\mb{Z}_2^T$ time-reversal is implemented anti-unitarily~\cite{wigner2012group}, the time-reversal operator can be written as a $T=UK$, where $K$ denotes complex conjugation in the basis with respect to which the MPS tensors $A^j$ are defined, and $U$ is a unitary satisfying $U\overline{U}=\pm\mbm{1}$ depending on whether time-reversal is implemented linearly or projectively (Section~\ref{subsec:TimeRev}). Hence, on the MPS tensors, a symmetry $g\in\msf{G}=\msf{H}\rtimes\mb{Z}_2^T$, $\msf{H}$ denoting the global on-site symmetry group, acts according to
\begin{equation}
	\sum_j \left(U_g\right)_{ij}C_g\left(A^j\right) = e^{i\varphi(g)}X_g^{-1}A^iX_g.
	\label{eq:TransformTensor2}
\end{equation}
The action of $C_g$ on $A^i$ is taking the complex conjugate only if $g$ contains a time-reversal.

From acting with time-reversal twice on the MPS tensor, it follows that in this case the matrices $X_g$ form a \emph{generalized} projective representation of the symmetry group, as their multiplication also contains the action $C_g$:
\begin{equation}
	X_gC_g\left(X_h\right) = e^{i\omega(g,h)}X_{gh},
\end{equation}
whereas the phases $\exp(i\varphi(g))$ obey
\begin{equation}
	e^{i\varphi(g)}C_g\left(e^{i\varphi(h)}\right) = e^{i\varphi(gh)}.
\end{equation}
Hence, $\varphi(g)$ and $\omega(g,h)$ satisfy the 1- and 2-cocycle constraints with a non-trivial group action that takes the conditioned complex conjugation $C_g$ into account. The 1-cocycles constraint reads
\begin{equation}
	\varphi(g) + \alpha^T_g\left(\varphi(h)\right) = \varphi(gh) \quad \mod 2\pi, \label{eq:Time1cocycle}
\end{equation}
whereas 2-cocycles satisfy
\begin{equation}
	\begin{multlined}
		\omega(g,h) + \omega(gh,k) =\\
		\qquad\qquad \omega(g,hk) + \beta^T_g\left(\omega(h,k)\right) \mod 2\pi, \label{eq:Time2cocycle}
	\end{multlined}
\end{equation}
where the group actions $\alpha^T_g$ and $\beta^T_g$ are multiplication by $-1$ if $g$ contains time-reversal. The SPT classification in this case is given by $H^1_{\alpha^T}\left(\msf{G}, \msf{U}_1\right)\times H^2_{\beta^T}\left(\msf{G}, \msf{U}_1\right)$. The equivalence classes of $H^2_{\beta^T}$ are given by $[\omega] = [\omega'] \iff \omega(g,h) = \omega'(g,h) + \gamma(g) + \beta_g^T\left(\gamma(h)\right) -\gamma(gh)$, whereas $H^1_{\alpha^T}$ classifies 1-cocycles up to equivalence of the form $\varphi(g)\mapsto\varphi(g) + \alpha_g^T(c) - c$ for an arbitrary constant $c\in[0,2\pi)$ as coboundaries of the form $\alpha_g^T(c) - c$ trivially solve the 1-cocycle condition (\ref{eq:Time1cocycle}).

In case the symmetry group contains global on-site symmetries combined with parity, $\msf{G}=\msf{H}\rtimes \mb{Z}_2^P$, we have that
\begin{equation}
	\sum_j \left(U_g\right)_{ij}T_g\left(A^j\right) = e^{i\varphi(g)}X_g^{-1}A^iX_g,
	\label{eq:TransformTensor3}
\end{equation}
where $T_g$ denotes taking the transpose if $g$ contains the parity transformation. Similarly as in the case of time-reversal, the gauge matrices multiply according to a generalized projective representation:
\begin{equation}
	X_gP_g\left(X_h\right) = e^{i\omega(g,h)}X_{gh},
	\label{eq:ProjRepPar}
\end{equation}
where $P_g$ is taking the inverse transpose if $g$ contains parity. Note, however, that if $g$ contains the parity transformation, $X_g$ can in general not be chosen unitary. As the conditioned transpose does not affect the phases $\varphi$, they form a one-dimensional linear representation. The SPT classification for $\msf{H}\rtimes \mb{Z}_2^P$ is thus in terms of $H^1\left(\msf{G}, \msf{U}_1\right)\times H^2_{\beta^P}\left(\msf{G}, \msf{U}_1\right)$. The 2-cocycles are given by
\begin{equation}
	\begin{multlined}
		\omega(g,h) + \omega(gh,k) =\\
		\qquad\qquad\omega(g,hk) + \beta^P_g\left(\omega(h,k)\right) \mod 2\pi,
		\label{eq:Parity2cocycle}
	\end{multlined}
\end{equation}
where, similarly as in the case of time-reversal, the group action $\beta^P_g$ is a multiplication with $- 1$ whenever $g$ contains a parity transformation. The equivalence classes in the second cohomology group are then $[\omega]=[\omega']\iff\omega(g,h) = \omega'(g,h) + \gamma(g) + \beta_g^{P}(\gamma(h)) -\gamma(gh)$.
\subsection{Computing group cohomology using the Smith normal form}
\label{sec:SmithNormal}
The problem of finding all (generalized) projective representations of a given finite symmetry group $\mathsf{G}$  or, equivalently, computing its second cohomology group $H^2_\beta\left(\msf{G}, \msf{U}_1\right)$ can be reduced to a problem in linear algebra that can be solved using the Smith normal form~\cite{yang2017}, as we now demonstrate. Our approach works for both trivial and non-trivial group actions $\beta$ and the same method can be used to compute other cohomology groups.

First note that the 2-cocycle equation (\ref{eq:2cocycle}) or its generalizations with non-trivial group actions (\ref{eq:Time2cocycle}, \ref{eq:Parity2cocycle}) can be written as the linear system
\begin{equation}
	\sum_j\Omega_{ij}^{(2,\beta)}\omega_j = 0, \quad \mod 2\pi,
	\label{eq:System}
\end{equation}
which has to be solved modulo $2\pi$. $\Omega^{(2,\beta)}$ is called the 2-coboundary map, where $\beta$ again refers to the group action. Every solution $\vec{\omega}$ to this linear system of equations constitutes a valid 2-cocycle.

Since $\Omega^{(2,\beta)}$ only has entries in $\mb{Z}$, which forms a principal ideal domain, $\Omega^{(2,\beta)}$ can be written in Smith normal form as follows~\cite{kaczynski2004}:
\begin{equation}
	P\Lambda R = \Omega^{(2,\beta)}.
\end{equation}
In this decomposition $P$ and $R$ are respectively $\left|\mathsf{G}\right|^3\times \left|\mathsf{G}\right|^3$ and $\left|\mathsf{G}\right|^2\times \left|\mathsf{G}\right|^2$ matrices that only contain integers and have determinant one (and thus have integer-valued inverses). $\Lambda$ also only contains integers, is $\left|\mathsf{G}\right|^3\times \left|\mathsf{G}\right|^2$-dimensional and is of the form
\begin{equation}
	\Lambda
	=
	\begin{pmatrix}
		\begin{array}{c|c}
			\text{diag}\left(d_1,d_2,...,d_r\right) & 0 \\
			\hline
			0                                       & 0
		\end{array}
	\end{pmatrix}
	,
\end{equation}
in which the non-zero elements $d_1,...,d_r$ along the diagonal, some of which might be one, are in increasing order, $d_1\leq d_2\leq...$, and every element is a divisor of the next, $d_i | d_{i+1}$. The Smith normal form $\Lambda$ is unique. Inserting this decomposition in the system of equations (\ref{eq:System}) gives rise to the solution
\begin{equation}
	\vec{\omega} = 2\pi R^{-1}\Lambda^+\vec{\nu}.
	\label{eq:Sol2Cocyc}
\end{equation}
$\Lambda^+$ denotes the (unique) Moore-Penrose pseudoinverse of $\Lambda$ that satisfies $\Lambda\Lambda^+\Lambda = \Lambda$ and which is found to be
\begin{equation}
	\Lambda^+
	=
	\begin{pmatrix}
		\begin{array}{c|c}
			\text{diag}\left(d_1^{-1},d_2^{-1},...,d_r^{-1}\right) & 0 \\
			\hline
			0                                                      & 0
		\end{array}
	\end{pmatrix}
	.
\end{equation}
$\vec{\nu}$ is an arbitrary vector that only contains integers.\par
Writing the solution (\ref{eq:Sol2Cocyc}) in components yields
\begin{equation}
	\omega_i = \sum_{j=1}^r 2\pi\left(R^{-1}\right)_{ij}\frac{\nu_j}{d_j}.
\end{equation}
Because $\vec{\nu}$ can be chosen freely, one can choose subsequently $\nu_i = \delta_{1,i}, \delta_{2,i},...,\delta_{r,i}$ to obtain a basis of the solution space that can be written as
\begin{equation}
	(\vec{\omega}_j)_i = \frac{2\pi}{d_j}\left(R^{-1}\right)_{ij}, \quad \forall j\in\{1,...,r\}.
	\label{eq:Sol2CocyBis}
\end{equation}
Hence, the 2-cocycles $\vec{\omega}$ are found to be the columns of $R^{-1}$. Since $R$ is full rank, all the solutions $\vec{\omega}$ are linearly independent. In particular, the non-trivial cocycles (below) can not be related by a coboundary, $\vec{\omega}'=\vec{\omega} + \Omega^{(1,\beta)}\vec{\varphi}$, where $\Omega^{(1,\beta)}$ denotes the 1-coboundary map and $\vec{\varphi}$ is a $|\msf{G}|$-dimensional vector containing arbitrary real numbers. To classify all possible solutions, we now consider the diagonal entries of $\Lambda$.

From (\ref{eq:Sol2CocyBis}) and the fact that $R^{-1}$ contains only integers, it follows that for every diagonal entry $d_i=1$, a trivial solution $\omega_i=0 \mod 2\pi$ is obtained.
The non-trivial solutions are those that correspond to entries $d_i>1$. From (\ref{eq:Sol2CocyBis}) and the fact that the solution space is $\mb{Z}$-linear, it follows that the cocycle $\vec{\omega}_j$ corresponding to some $d_j$ generates a cyclic group. Now note that not all elements of $\vec{\omega}_j$ can be divisible by $d_j$ (or any of its prime factors) as this would be in contradiction with the fact that $R^{-1}$ has determinant one. Hence, the cyclic group generated by $\vec{\omega}_j$ is $\mb{Z}_{d_j}$.
Finally, the zero entries of $\Lambda$ can also be discarded in the cohomology as these correspond to trivial solutions of the cocycle equation that can be multiplied by arbitrary phases and thus correspond to coboundaries.

In conclusion following picture arises. Given some group $\msf{G}$ one can write down the coboundary map $\Omega_{ij}^{(2,\beta)}$ that can be brought in Smith normal form $P\Lambda R$. The diagonal entries of $\Lambda$, $d_1,d_2,...,d_r$, determine the second cohomology group which is then of the form $\mb{Z}_{d_1}\times\mb{Z}_{d_2}\times ... \times \mb{Z}_{d_r}$, with the understanding that $\mb{Z}_1=\{e\}$ denotes the trivial group and that all zero diagonal entries can be discarded. The non-trivial 2-cocycles in some arbitrary gauge correspond then to the columns of $R^{-1}$, weighted by the appropriate factor $2\pi/d_j$.
\subsection{Zero correlation length SPT ansatz}
\label{sec:ansatz}
Given some SPT phase characterized by $\left(\left[\varphi\right],\left[\omega\right]\right)\in  H^1_\alpha\left(\msf{G}, \msf{U}_1\right) \times  H^2_\beta\left(\msf{G}, \msf{U}_1\right)$, it is possible to explicitly construct zero-correlation-length MPS tensors that transform according to (\ref{eq:TransformTensor}), (\ref{eq:TransformTensor2}) or (\ref{eq:TransformTensor3}).
Firstly, a projective representation $X_g$ in the class $\left[\omega\right]$ and with trivial 1-cocycle can be constructed with virtual dimension $\left|\msf{G}\right|$, by taking the $X_g$ to form the $\omega$-projective regular representation of $\msf{G}$ (\ref{eq:ProjRep}). The dimension of the local physical Hilbert space is then $\left|\msf{G}\right|^2$. Concretely, this representative state and the regular representation are given by:
\begin{numcases}{}
	\left(X_g\right)_{g_1,g_2} = \delta_{g_1, g g_2} e^{i\omega(g, g_2)} \label{eq:ProjRep}\\
	\left(V^{h_1,h_2}\right)_{g_1,g_2} = \frac{1}{\left|\msf{G}\right|} \delta_{h_1,g_1} \delta_{h_2, g_2} e^{i\omega(h_2,h_2^{-1} h_1)}.
	\label{eq:FixedPoint}
\end{numcases}
Furthermore, the physical group action $U_g$ is given by $L_g \otimes L_g$, with $L_g$ the linear left regular representation. Hence, $(U_g)_{k_1k_2,h_1h_2} = \delta_{k_1,gh_1} \delta_{k_2,gh_2}$.
Pictorially:
\begin{equation}
	\includegraphics[valign=c]{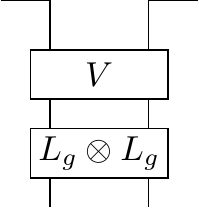}
	=
	\includegraphics[valign=c]{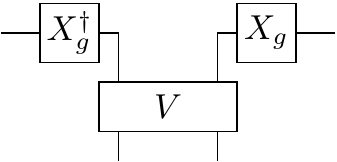}
	.
\end{equation}
Note that the Kronecker deltas in the definition of the MPS tensor indicate its valence bond structure, which is modified only by a unitary diagonal transformation. Hence, it follows readily that the transfer matrix $\mb{E} = \sum_{h_1,h_2} \left(V^{h_1,h_2}\right)_{g_1,g_2} \otimes \overline{\left(V^{h_1,h_2}\right)}_{g_1',g_2'}$ is idempotent, $\mb{E}^2=\mb{E}$, implying that the ansatz (\ref{eq:FixedPoint}) has zero correlation length and thus defines an RG fixed point~\cite{Verstraete2005}.\\

Similarly, in case of a non-trivial 1-cocycle $\varphi$, the ansatz is readily modified to also include this cocycle by adding diagonal matrices to the two physical legs:
\begin{equation}
	\includegraphics[valign=c]{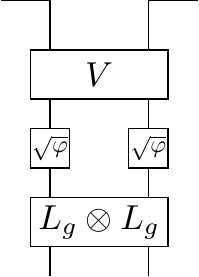}
	=
	e^{i\varphi(g)}
	\hspace{-4pt}
	\left[
	\includegraphics[valign=c]{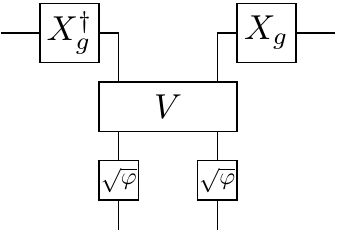}
	\right],
\end{equation}
where $(\sqrt{\varphi})_{g_1,g_2}=\delta_{g_1,g_2}e^{\frac{i}{2}\varphi(g)}$.\par
The ansatz can also capture the case of time-reversal and parity symmetry. In the first case we take the physical action to be $L_g\otimes L_g$ combined with the conditioned complex conjugation:
\begin{equation}
	\hspace{-4pt}
	\includegraphics[valign=c]{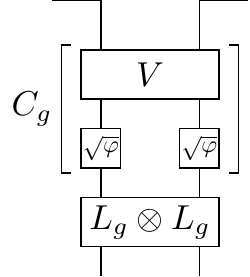}
	=
	e^{i\varphi(g)}
	\hspace{-4pt}
	\left[
	\includegraphics[valign=c]{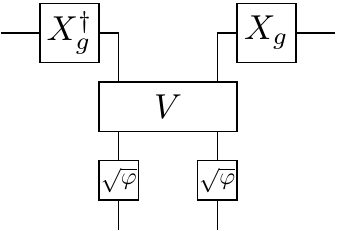}
	\right],
\end{equation}
were we still have $\left(X_g\right)_{g_1,g_2} = \delta_{g_1, g g_2} e^{i\omega(g, g_2)}$.\par
The ansatz in case of parity symmetry and parity+time-reversal requires following conditioned `swap' tensor
\begin{equation}
	\includegraphics[valign=c]{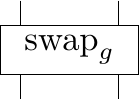}
	=
	\begin{cases}
		\includegraphics[valign=c]{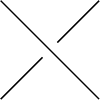}\ ,\quad \text{$g$ contains parity,} \\
		\\
		\includegraphics[valign=c]{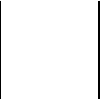}\ ,\quad \text{else.}
	\end{cases}
\end{equation}
The ansatz in this case then amounts to
\begin{equation}
	\hspace{-4pt}
	\includegraphics[valign=c]{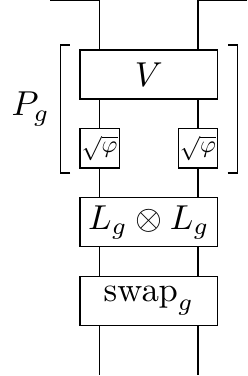}
	=
	e^{i\varphi(g)}
	\hspace{-4pt}
	\left[
	\includegraphics[valign=c]{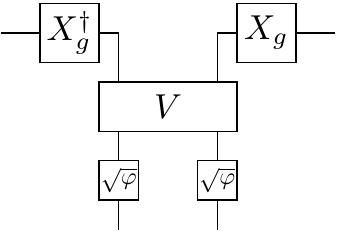}
	\right],
\end{equation}
where $P_g$ acts according to
\begin{equation}
	\includegraphics[valign=c]{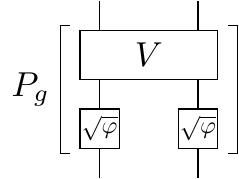}
	=
	\includegraphics[valign=c]{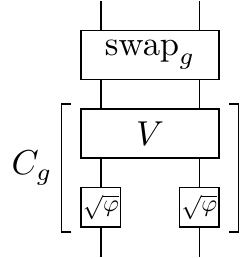}
	.
\end{equation}
\subsubsection{Example: \texorpdfstring{$\mb{Z}_2\times\mb{Z}_2$}{Z2xZ2}}
\label{sec:Z2Z2}
The smallest finite group with a non-trivial 2-cocycle is $\mb{Z}_2\times\mb{Z}_2$ as $H^2\left(\mb{Z}_2\times\mb{Z}_2,\msf{U}_1\right)=\mb{Z}_2$. $\mb{Z}_2\times\mb{Z}_2$ is exactly the on-site symmetry group of the AKLT model and the cluster state~\cite{affleck1987, raussendorf2001, briegel2001, Verstraete2005}. The latter admits a description as an injective bond dimension two MPS~\cite{verstraete2004}:
\begin{numcases}{}
	A^0 = |+)(0| \\
	A^1 = |-)(1|
	,
\end{numcases}
where $|\pm)=\frac{1}{\sqrt{2}}(|0)\pm|1))$. This MPS description can be derived from the fact that the cluster state is obtained starting from the product state $\ket{+}^{\otimes N}$, acting on pairs of neighbouring spins with controlled Z gates and projecting onto the physical degrees of freedom with the projector $\ketbra{0}{00} + \ketbra{1}{11}$. The cluster state is no RG fixed point but the fixed point is obtained after blocking only two sites ~\cite{Verstraete2005}. The MPS description of this RG fixed point is then given by
\begin{numcases}{}
	A^{00} = \frac{1}{\sqrt{2}}|+)(0| \\
	A^{01} = \frac{1}{\sqrt{2}}|+)(1| \\
	A^{10} = \frac{1}{\sqrt{2}}|-)(0| \\
	A^{11} = \frac{-1}{\sqrt{2}}|-)(1|
	\label{eq:RGFixedPoint}
	.
\end{numcases}
The normalization of the state is chosen in such a way that the unique non-zero eigenvalue of the transfer matrix is 1. This state is in the non-trivial SPT class with on-site $\mb{Z}_2\times\mb{Z}_2$ symmetry. $\mb{Z}_2\times\mb{Z}_2$ acts linearly on the physical level as $S((0,1))=\sigma_X\otimes\sigma_X$, $S((1,0))=\mbm{1}\otimes\sigma_X$, and the symmetry is represented projectively on the virtual level by $X((0,1))=i\sigma_Y$, $X((1,0))=\sigma_X$, $X((1,1))=\sigma_Z$.

Since our correlation length zero ansatz (\ref{eq:FixedPoint}) that is constructed from a given non-trivial 2-cocycle of $\mb{Z}_2\times\mb{Z}_2$ is in the same SPT phase as the cluster state, we can expect that our ansatz is up to a basis transformation and gauge transformations the product of the cluster state fixed point and a trivial state. We now show that our ansatz indeed reduces to the product of the RG fixed point cluster state dressed with a trivial dimer state and demonstrate how to construct this basis transformation and gauge transformation explicitly.

The linear regular representation of $\mb{Z}_2\times\mb{Z}_2$ acting on the physical level of our ansatz reads
\begin{numcases}{}
	L_{(0,1)} = \mbm{1}\otimes\sigma_X \\
	L_{(1,0)} = \sigma_X\otimes\mbm{1} \\
	L_{(1,1)} = \sigma_X\otimes\sigma_X.
\end{numcases}
A non-trivial 2-cocycle of $\mb{Z}_2\times\mb{Z}_2$ in a particular gauge is given by
\begin{align}
	\omega(g,h) & =\bordermatrix{
		& (0,0) & (1,0) & (0,1) & (1,1)\cr
		(0,0) & 1 & 1 & 1 & 1 \cr
		(1,0) & 1 & 1 & 1 & 1 \cr
		(0,1) & 1 & -1 & -1 & 1 \cr
		(1,1) & 1 & -1 & -1 & 1 \cr
	},
\end{align}
where $g$ labels the rows. The projective regular representation corresponding to this cocycle (\ref{eq:ProjRep}) reduces to
\begin{numcases}{}
	U^\dagger X_{(0,1)}U = \mbm{1}\otimes i\sigma_Y\label{eq:BlockProjRep1}\\
	U^\dagger X_{(1,0)}U = \mbm{1}\otimes -\sigma_X\\
	U^\dagger X_{(1,1)}U = \mbm{1}\otimes \sigma_Z,\label{eq:BlockProjRep3}
\end{numcases}
in the basis $U$ given by
\begin{equation}
	U = \frac{1}{\sqrt{2}}
	\begin{pmatrix}
		0  & 1  & 1 & 0  \\
		1  & 0  & 0 & -1 \\
		-1 & 0  & 0 & -1 \\
		0  & -1 & 1 & 0
	\end{pmatrix}.
\end{equation}
We then take this unitary $U$ as a gauge transformation of our representative MPS ansatz. This gauge transformation then intertwines between the projective regular representation of $\mb{Z}_2\times\mb{Z}_2$ and the block diagonal projective representation given in (\ref{eq:BlockProjRep1}-\ref{eq:BlockProjRep3}). Similarly as in the case of linear representation theory, this illustrates how the projective regular representation of $\mb{Z}_2\times\mb{Z}_2$ falls apart in projective irreps, given by the Pauli matrices, where each irrep appears with multiplicity equal to its dimension~\cite{Cheng}.

The physical basis transformation $W$ that brings the representative MPS $V$ in the form of the cluster state fixed point dressed with a trivial dimer state,
\begin{equation}
	\includegraphics[valign=c]{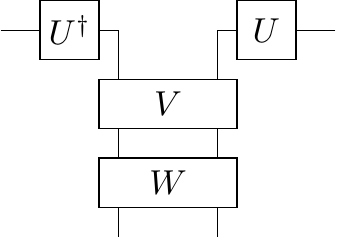}
	=
	\includegraphics[valign=c]{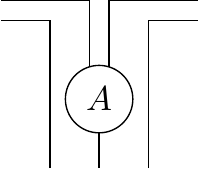},
\end{equation}
is then immediately found by considering the QR decomposition of our ansatz and the dressed cluster state interpreted as matrix from the physical to virtual level, $QR=V(\overline{U}\otimes U)$, $\widetilde{Q}DD^\dagger\widetilde{R}=\mbm{1}\otimes A\otimes\mbm{1}$, where $D$ is a diagonal matrix containing only phases that fixes the gauge freedom of the QR decomposition in such a way that $D^\dagger\widetilde{R}=R$. The basis transformation then reads $W=\widetilde{Q}DQ^\dagger$. It can then be checked that this $W$ intertwines the physical representation of the $\mb{Z}_2\times\mb{Z}_2$ symmetry group acting on the dressed cluster state and our ansatz:
\begin{numcases}{}
	W^\dagger (\mbm{1}\otimes\sigma_X\otimes\sigma_X\otimes\mbm{1}) W = L_{(0,1)} \otimes L_{(0,1)}\\
	W^\dagger (\mbm{1}\otimes\mbm{1}\otimes\sigma_X\otimes\mbm{1}) W = L_{(1,0)} \otimes L_{(1,0)}\\
	W^\dagger (\mbm{1}\otimes\sigma_X\otimes\mbm{1}\otimes\mbm{1}) W = L_{(1,1)} \otimes L_{(1,1)}.
\end{numcases}
Notice that in this example we were required to choose the symmetry group of the dimer to be trivial to match the symmetry of our ansatz, even though the dimer has the full $\msf{U}_2$ symmetry.
\subsection{A generalized ansatz}
\label{sec:GenAnsatz}
As we have demonstrated, given a 1- and 2-cocycle of some finite symmetry group $\msf{G}$, one can explicitly construct a bond dimension $|\msf{G}|$ injective correlation length zero MPS that belongs to the SPT class corresponding to the cohomology classes represented by these cocycles. However, as shown for the explicit example of $\mb{Z}_2\times\mb{Z}_2$, our ansatz could be written as a cluster state fixed point and a completely disentangled trivial dimer state that doubles the dimension of the virtual Hilbert space. This redundancy is a consequence of the fact that the physical symmetry action is fixed as being the tensor product of two regular representations of $\msf{G}$. Hence, the question rises if an MPS can be constructed for any given representation of the symmetry on the physical level such that the bond dimension is as small as possible. This can in principle be done as follows. We restrict to the case of unitary on-site symmetries.

We first note that every projective representation of a finite group $\msf{G}$ can be lifted to a linear representation of a larger finite covering group $\widetilde{\msf{G}}$~\cite{CurtisReiner}. This covering group fits in following central exact sequence
\begin{equation}
	1 \longrightarrow \msf{A} \longrightarrow \widetilde{\msf{G}} \longrightarrow \msf{G} \longrightarrow 1,
\end{equation}
where $\msf{A}=H^2\left(\msf{G},\msf{U}_1\right)\leq Z(\widetilde{\msf{G}})$~\cite{Bessenrodt1994}. It should be noted that the covering group is generically not unique. Consider for example the case of $\mb{Z}_2\times\mb{Z}_2$. Two distinct covering groups of $\mb{Z}_2\times\mb{Z}_2$ are the dihedral group $\msf{D}_4=\expval{r,s|r^4=s^2=(rs)^2=1}$ and the quaternion group $\msf{Q}=\expval{a,b|a^4=1,a^2=b^2, b^{-1}ab=a^{-1}}$, both of which are of order 8. Indeed, the two-dimensional projective irrep of $\mb{Z}_2\times\mb{Z}_2$ corresponding to the non-trivial class of $H^2\left(\mb{Z}_2\times\mb{Z}_2,\msf{U}_1\right)=\mb{Z}_2$ given by $\{\sigma_X,i\sigma_Y,\sigma_Z\}$ is lifted to the faithful two-dimensional irrep of $\msf{D}_4$ by taking $i\sigma_Y=r$ and $\sigma_Z=s$ as generators, whereas a gauge transformation of this projective representation, $\{\sigma_X,i\sigma_Y,\sigma_Z\}\mapsto\{i\sigma_X,i\sigma_Y,i\sigma_Z\}$, yields an equivalent projective representation which is lifted to the faithful two-dimensional irrep of $\msf{Q}$ by identifying the generators of the quaternion group as $a=i\sigma_Y$, $b=i\sigma_Z$.

The classification and construction of the projective irreps of a finite group thus reduces in this way to the linear representation theory of its covering group. To construct the aforementioned MPS that transforms according to a given physical representation $\Pi$ of the symmetry group $\msf{G}$, one chooses the smallest irrep $\Gamma$ of $\widetilde{\msf{G}}$, which projects down to a projective irrep of $\msf{G}$ belonging to a certain cohomology class $[\omega]$, such that $\Pi$ is contained in the tensor product $\overline{\Gamma}\otimes\Gamma$. The explicit MPS tensor is then chosen as the projector of $\overline{\Gamma}\otimes\Gamma$ on the $\Pi$-sector, and the virtual symmetry action is $\Gamma$.

Consider as example again $\mb{Z}_2\times\mb{Z}_2$ and its covering group $\msf{Q}$. The irreps of $\msf{Q}$ are the trivial representation $1$, three non-trivial one-dimensional sign representations, $\Gamma_1,\Gamma_2,\Gamma_3$, and the two-dimensional faithful representation given by the Pauli matrices, $\Delta$. Choosing then a sign representation of $\mb{Z}_2\times\mb{Z}_2$ as physical symmetry, this representation can be lifted to a one-dimensional sign representation $\Gamma_i$ of $\msf{Q}$. From the fact that $\Delta\otimes\Delta\cong 1\oplus\Gamma_1\oplus \Gamma_2\oplus \Gamma_3$, it follows that the MPS with the smallest bond dimension which transforms according to a sign representation of $\mb{Z}_2\times\mb{Z}_2$ is exactly one that projects $\Delta\otimes\Delta$ onto $\Gamma_i$. Since the virtual representation $\Delta$ is exactly the projective irrep of $\mb{Z}_2\times\mb{Z}_2$, one can immediately conclude that such a state is in the same SPT phase as the cluster state.

This construction also applies to the case of Lie groups. Consider for example the symmetry group $\msf{SO}_3$. Choosing the physical symmetry representation to be the {\bf 1} of $\msf{SO}_3$, one can choose the virtual representation to be the $\bf \frac{1}{2}$. The projector of $\bf \frac{1}{2}\otimes\frac{1}{2}\cong 0\oplus 1$ on the $\bf 1$ subspace then exactly results in the MPS description of the AKLT state which belongs to the non-trivial SPT class of $H^2\left(\msf{SO}_3,\msf{U}_1\right)=\mb{Z}_2$.

Notice that this constructing generically does not give rise to correlation length zero states due to the projection on the correct physical symmetry sector.

\section{Frieze symmetric MPS}
\label{sec:frieze}

\emph{In this section we derive the SPT classification of MPS invariant under frieze symmetries. We do so from starting from a general injective MPS and invoking the symmetry, which ultimately leads to topological indices that cannot be changed by a symmetry-preserving constant depth quantum circuit.}

\Sep

We introduce following notation:
\begin{equation}
	\mbm{1}_{n,m} \coloneqq
	\mbm{1}_{n} \oplus -\mbm{1}_m
	.
\end{equation}
This matrix satisfies $\mbm{1}_{n,m}=\mbm{1}_{n,m}^\top=\mbm{1}_{n,m}^{-1}$. Furthermore:
\begin{equation}
	\Gamma
	\coloneqq
	\mbm{1}_n\otimes
	\begin{pmatrix}
		0  & 1 \\
		-1 & 0
	\end{pmatrix}
	= -\Gamma^\top
	= - \Gamma^{-1}
	.
\end{equation}
And finally:
\begin{equation}
	\sqrt{\Gamma} = \mbm{1}_n\otimes
	\frac{1}{\sqrt{2}}
	\begin{pmatrix}
		1  & 1 \\
		-1 & 1
	\end{pmatrix},\
	\sqrt{\Gamma}^\top
	= \sqrt{\Gamma}^{-1}.
\end{equation}
$\sim\mathbf{F_0}\sim$ As mentioned in the introduction, every translationally invariant MPS can be brought in a uniform form by an appropriate gauge transformation~\cite{perezgarcia2007}. There are no topological obstructions to do so and hence there are no non-trivial SPT phases.

$\sim\mathbf{F_V}\sim$ Reflection around the horizontal axis can be thought of as an internal $\mb{Z}_2$ transformation of the MPS tensors. In order to impose such a reflection, we consider a uniform MPS with two physical legs that mimic an internal structure of the local degrees of freedom,
\begin{equation}
	\includegraphics[valign=c]{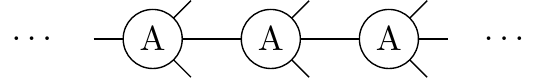}
	.
	\label{eq:TwoLeggeAnsatz}
\end{equation}
Under the reflection the tensor $A$ transforms according to
\begin{equation}
	A^{ji} = e^{i\theta}X^{-1}A^{ij}X.
\end{equation}
After applying this symmetry transformation twice we obtain
\begin{equation}
	A^{ij} = e^{2i\theta}X^{-2}A^{ij}X^{2}.
\end{equation}
From the fundamental theorem it follows that $\theta$ is a topological index, $\theta = 0,\pi \mod 2\pi$, and $X^2=e^{i\chi}\mbm{1}$, however this phase can be absorbed in $X$ such that $X$ squares to the identity. $F_V$ can thus protect one non-trivial SPT phase characterized by $\theta=\pi$.

If the MPS is in left canonical form, it can be shown that $X$ is unitary, which together with $X^2=\mbm{1}$ implies that $X$ can be written as $X=U^\dagger\mbm{1}_{n,m}U$ for some unitary $U$ and $n+m=D$. With the gauge transformation $A^{ij}\mapsto\tilde{A}^{ij} = UA^{ij}U^\dagger$ the MPS tensors $\tilde{A}$ have definite $V$-parity:
\begin{equation}
	\tilde{A}^{ji} = \pm \mbm{1}_{n,m}\tilde{A}^{ij}\mbm{1}_{n,m},
	\label{eq:VParity}
\end{equation}
thus tremendously reducing the number of variational degrees of freedom. The signature of $\mbm{1}_{n,m}$ is irrelevant for the classification of SPT orders. Indeed, notice that the $F_V$ symmetry can be thought of as an on-site $\mb{Z}_2$ symmetry acting on the physical level that is implemented by $\sum_{(j_1,j_2)}\left(U_1\right)^{(i_1,i_2)}_{(j_1,j_2)}A^{j_1,j_2}$ with $\left(U_1\right)^{(i_1,i_2)}_{(j_1,j_2)} = \delta_{j_2}^{i_1}\delta_{j_1}^{i_2}$ and $1$ denoting the non-trivial element in $\mb{Z}_2$. This unitary acting on the physical level than translates to the $\mbm{1}_{n,m}$, also a (reducible) representation of $\mb{Z}_2$, acting on the virtual level.  It was shown in~\cite{Chen2011a, Chen2011b} that every such representation of $\mb{Z}_2$ at the virtual level gives rise to the same SPT phase because the $\mbm{1}_{k,l}$ for all $k,\ l$ belong to the same (trivial) second cohomology class of  $H^2\left(\mb{Z}_2,\msf{U}_1\right)=\left\{e\right\}$. In particular, the MPS is in the same phase as the MPS that transforms according to
\begin{equation}
	\tilde{A}^{ji} = \pm \tilde{A}^{ij}.
\end{equation}
This result fits within the well understood SPT classification in terms of group cohomology for the case of an on-site $\mathbb{Z}_2$ symmetry~\cite{Chen2010,Chen2011a,Chen2011b,Chen2013,Schuch2011}. Indeed, we could have implemented the vertical reflection more generally as
\begin{equation}
	A^i \mapsto \sum_j R_{ij}A^j,
\end{equation}
where $R_{ij}$ is a (linear) representation of $\mb{Z}_2$, $R^2=\mbm{1}$. As mentioned above $H^2\left(\mb{Z}_2,\msf{U}_1\right)=\left\{e\right\}$ meaning there are no non-trivial projective representations of $\mb{Z}_2$ that can protect the phase; however $H^1\left(\mb{Z}_2,\msf{U}_1\right)=\mb{Z}_2$, which is reflected in the fact that $e^{i\theta}=\pm 1$ and as such can be understood as the distinction between $V$-odd/even tensors $\tilde{A}^{ij}$.

$\sim\mathbf{F_H}\sim$ $F_H$ contains apart from a translation generator a reflection in the horizontal direction and as such corresponds to the parity transformation previously considered in~\cite{Chen2011a}. Without loss of generality we can start from a uniform MPS (\ref{eq:UniformMPS}) generated by a tensor $A^i$. Imposing the $\mb{Z}_2$ $H$-parity symmetry, we can write
\begin{equation}
	\sum_j R_{ij} \left(A^j\right)^\top = e^{i\theta} X^{-1}A^iX,
	\label{eq:HParityOnce}
\end{equation}
$R$ being an involutory unitary matrix. Doing the transformation twice results in
\begin{equation}
	A^i = e^{2i\theta}X^{\top}X^{-1}A^iXX^{-\top},
\end{equation}
from which it follows that
\begin{equation}
	\begin{cases}
		\theta = 0, \pi \quad \mod 2\pi \\
		X^\top = e^{i\chi}X.
		\label{eq:CondF_H}
	\end{cases}
\end{equation}
From the last equation we conclude that $\chi=0, \pi \mod 2\pi$. Together with the first equation, this leads to a $\mb{Z}_2\times\mb{Z}_2$ classification. We now show that every $F_H$ symmetric MPS can be brought in a form in which the local tensors are (skew-)symmetric under $H$-parity, possibly at the cost of introducing extra bond tensors~\cite{jiang2015}.

First consider the case that $X$ is symmetric. Using the Autonne-Takagi decomposition, we can write $X=CC^\top$ for some complex matrix $C$. From substitution in (\ref{eq:HParityOnce}) it immediately follows that defining $A^i \mapsto \tilde{A}^i=C^{-1}A^iC$ yields a uniform MPS generated by $\tilde{A}^i$ in which the tensors $\tilde{A}^i$ have a definite $H$-parity given by the topological index $\theta$.

Now assume that $X$ is skew-symmetric, which together with its invertibility requires that $D=2n$ is even. In that case we similarly can write $X=C\Gamma C^\top$ using the Youla normal form~\cite{youla1961normal}. Again substituting this in (\ref{eq:HParityOnce}), we can identify $A^i \mapsto \tilde{A}^i=\sqrt{\Gamma}C^{-1}A^iC\sqrt{\Gamma}$ as the transformation to construct local tensors with a well-defined $H$-parity. However, this transformation is not an actual gauge transformation in the aforementioned sense since, in order to bring the MPS in the desired form, we should insert $\mbm{1}=C\sqrt{\Gamma}\Gamma^\top\sqrt{\Gamma}C^{-1}$ between the $A$-tensors, which after making the transformation $A^i\mapsto\tilde{A}^i$ leaves us with residual bond tensors $\Gamma^\top$:
\begin{equation}
	\includegraphics[valign=c]{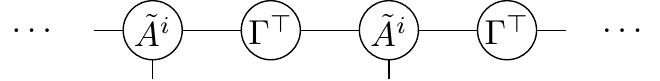}
\end{equation}
The SPT classification for $F_H$ corresponds to $H^1\left(\mb{Z}_2^P,\msf{U}_1\right)\times H^2_{\beta^P}\left(\mb{Z}_2^P,\msf{U}_1\right)=\mb{Z}_2\times \mb{Z}_2$, where the first $\mb{Z}_2$ factor corresponds to $\theta=0,\pi$ and the second $\mb{Z}_2$ corresponds to the phase appearing in the generalized projective representation $X$ of the parity symmetry group $\mb{Z}_2^P$.

$\sim\mathbf{F_G}\sim$ The $F_G$ symmetry group contains a generator of translations $T$ and a glide reflection $G$ which are related through $G^2=T$. Therefore, we will consider an MPS ansatz which is translationally invariant under shifts over two sites. The glide reflection will then be implemented as a shift over one site followed by a reflection around the horizontal axis. Hence, without loss of generality we can take this MPS ansatz to be
\begin{equation}
	\includegraphics[valign=c]{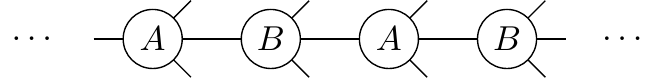}
	\label{eq:TwoSiteAnsatz}
\end{equation}
Invariance under glide reflections relates the tensors $A$ and $B$ up to a gauge transformation which can in general be different on the $A-B$- and $B-A$-bonds:
\begin{equation}
	\begin{cases}
		A^{ji} & = e^{i\theta_A}X^{-1}B^{ij}Y \\
		B^{ji} & = e^{i\theta_B}Y^{-1}A^{ij}X
	\end{cases}
	.
	\label{eq:glide}
\end{equation}
Carrying out this transformation twice then results in
\begin{equation}
	\begin{cases}
		A^{ij} & = e^{i\left(\theta_A + \theta_B\right)} X^{-1}Y^{-1} A^{ij} XY \\
		B^{ij} & = e^{i\left(\theta_A + \theta_B\right)} Y^{-1}X^{-1} B^{ij} YX
	\end{cases}
	.
	\label{eq:GlideTwice}
\end{equation}
Blocking two sites and using the fundamental theorem of MPS yields following conditions on the phases and gauge matrices:
\begin{equation}
	\begin{cases}
		2\left(\theta_A + \theta_B\right) = 0 \quad \mod 2\pi \\
		X = e^{i\chi} Y^{-1}
	\end{cases}
	.
\end{equation}
By absorbing a phase factor $e^{-i\frac{\chi}{2}}$ in both $X$ and $Y$, $X$ and $Y$ are each other inverses. Substituting $2\left(\theta_A + \theta_B\right) = 0 \ \mod 2\pi$ in (\ref{eq:GlideTwice}) shows that only the case $\theta_A + \theta_B = 0 \ \mod 2\pi$ can survive. We are free to choose eg. $\theta_A=0$. From this it then follows that $B^{ji}=XA^{ij}X$ such that after redefining $A^{ij}\mapsto\tilde{A}^{ij}=A^{ij}X$ the MPS can be brought in following canonical form:
\begin{equation}
	\includegraphics[valign=c]{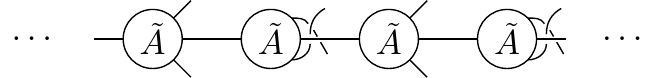}
	\label{eq:GlideMPS}
\end{equation}
In case of glide reflection symmetry there are thus no non-trivial SPT phases. This is again in line with the cohomological classification. Since the glide reflection should really be thought of as a generelized, dressed translation operator, the group cohomology classifying the SPT classes of glide reflection symmetry is that of the trivial group, which is trivial~\footnote{There are two distinct consistent ways in which a two-site unit cell structure can be compatible with a  $\mb{Z}_2$ symmetry action. These are classified by $H^1(\mb{Z}_2,\mb{Z}_2)=\mb{Z}_2$, which can be computed by hand. The two distinct 1-cocycles exactly correspond to a local $\mb{Z}_2$ symmetry and to a glide reflection.}.

$\sim\mathbf{F_R}\sim$ Starting from a two-legged uniform MPS ansatz (\ref{eq:TwoLeggeAnsatz}), we impose the rotation symmetry as
\begin{equation}
	\left(A^{ji}\right)^\top = e^{i\theta}X^{-1}A^{ij}X.
\end{equation}

Applying this symmetry twice and using the fundamental theorem, it immediately follows that $\theta=0,\pi \mod 2\pi$ and $X^\top=e^{i\chi}X$, exactly what was found in case of $F_H$ symmetry. The canonical form of an $F_R$ invariant MPS is thus again one in which the tensors have definite $R$-parity, again at the cost of introducing extra bond tensors between neighboring sites if $X$ is skew-symmetric.

$\sim\mathbf{F_{RG}}\sim$ Different $F_{RG}$-symmetric SPT phases ought to be classified by $H^1\left(\mb{Z}_2^P,\msf{U}_1\right)\times H^2_{\beta^P}\left(\mb{Z}_2^P,\msf{U}_1\right)=\mb{Z}_2\times \mb{Z}_2$, as again the glide reflection should be thought of as a generalized translation. To obtain this classification we first impose glide reflection as in (\ref{eq:glide}) on the ansatz (\ref{eq:TwoSiteAnsatz}), and again we find that under glide reflection the two tensors transform according to
\begin{equation}
	\begin{cases}
		A^{ji} &= XB^{ij}X \\
		B^{ji} &= X^{-1}A^{ij}X^{-1}
	\end{cases}.
\end{equation}
We obtain no topological indices from the glide reflection symmetry alone.

The rotation now acts on the tensors as
\begin{equation}
	\begin{cases}
		(A^{ji})^\top &= e^{i\psi_A}W^{-1}B^{ij}Y \\
		(B^{ji})^\top &= e^{i\psi_B}Y^{-1}A^{ij}W
	\end{cases},
\end{equation}
where the reflection center lies on an $A-B$-bond.

We then impose $R^2=1$ and $RGRG=1$, and after some lengthy algebra we ultimately find the transformation rules
\begin{equation}
	\begin{cases}
		(A^{ji})^\top &= e^{-i\sigma}X^\top Y^{-1}XB^{ij}Y\\
		(B^{ji})^\top &= e^{i\sigma}Y^{-1}A^{ij}X^{-1}YX^{-\top}\\
		Y^\top &= e^{i\chi}Y
	\end{cases}
	,
\end{equation}
where $\chi,\sigma = 0,\pi \mod 2\pi$, hence giving rise to the 3 anticipated non-trivial SPT phases. Here, the phase $\chi$ can be identified with the $H^2_{\beta^P}\left(\mb{Z}_2^P,\msf{U}_1\right)=\mb{Z}_2$, whereas $\sigma$ corresponds to $H^1\left(\mb{Z}_2^P,\msf{U}_1\right)$. It should be noted that the topological index $\chi$ arises purely from the rotational $\mb{Z}_2^P$ symmetry and that $\sigma$ finds its origin in the non-trivial constraint $RGRG=1$ interlocking the glide reflection and the rotation symmetry.

$\sim\mathbf{F_{VH}}\sim$ We consider again the two-legged uniform ansatz (\ref{eq:TwoLeggeAnsatz}). First imposing reflection around the horizontal axis yields
\begin{equation}
	A^{ji} = e^{i\theta_V}X^{-1}A^{ij}X.
	\label{eq:Vrefl}
\end{equation}
Carrying out this symmetry operation twice results in the same conditions on $\theta_V$ and $X$ as in the case of $F_V$: $\theta_V = 0,\pi \mod 2\pi$, $X^2=\mbm{1}$. $X$ is again unitary and again we can write $X=U^\dagger\mbm{1}_{n,m}U$ for some unitary $U$. As was explained, the signature of $\mbm{1}_{n,m}$ is irrelevant.\par
Imposing the second reflection, one obtains
\begin{equation}
	\left(A^{ij}\right)^\top = e^{i\theta_H}Y^{-1}A^{ij}Y,
	\label{eq:Hrefl}
\end{equation}
which implies that $\theta_H=0,\pi \mod 2\pi$, $Y=\pm Y^\top$. Finally we impose that the horizontal and vertical reflection commute on the physical level. Using (\ref{eq:Vrefl}) and (\ref{eq:Hrefl}), we conclude that $YX^{-\top}=e^{i\psi}XY$. Using $X^2=\mbm{1}$, it follows that $\psi=0,\pi$. In this way a $\mb{Z}_2^{\times 2}\times\mb{Z}_2^{\times 2}$ classification is obtained, exactly in line with the cohomological classification $H^1\left(\mb{Z}_2\rtimes \mb{Z}_2^P,\msf{U}_1\right)\times H^2_{\beta^P}\left(\mb{Z}_2\rtimes\mb{Z}_2^P,\msf{U}_1\right)=\mb{Z}_2^{\times 2}\times\mb{Z}_2^{\times 2}$.
The first factors $\mb{Z}_2^{\times 2}$ originating from the first cohomology group correspond to $\theta_V$ and $\theta_H$ and the last factors $\mb{Z}_2^{\times 2}$ correspond to the gauge matrix $Y$ being (anti-)symmetric and the generalized (anti-)commutation relation among $X$ and $Y$, $YX^{-\top}=\pm XY$. The correspondance between the 2-cocycles of $H^2_{\beta^P}\left(\mb{Z}_2\rtimes\mb{Z}_2^P,\msf{U}_1\right)=\mb{Z}_2^{\times 2}$ and the topological indices obtained from the MPS picture can be made a bit more explicit as follows. One can show that there always exists a gauge in which the cocycles of $H^2_{\beta^P}\left(\mb{Z}_2\rtimes\mb{Z}_2^P,\msf{U}_1\right)=\mb{Z}_2^{\times 2}$ are of the form
\begin{align}
	\omega(g,h) & =\bordermatrix{
		& (0,0) & (1,0) & (0,1) & (1,1)\cr
		(0,0) & 1 & 1 & 1 & 1 \cr
		(1,0) & 1 & 1 & 1 & 1 \cr
		(0,1) & 1 & \zeta _1 & \zeta _2 & \zeta _1 \zeta _2 \cr
		(1,1) & 1 & \zeta _1 & \zeta _2 & \zeta _1 \zeta _2 \cr
	},\label{eq:CohomologyZ2Z2P}
\end{align}
where rows are labelled by $g$ and $\zeta_1,\zeta_2=\pm 1$.
These signs $\zeta_1,\zeta_2$ then exactly correspond to the topological indices obtained from the MPS computation above since it follows from (\ref{eq:ProjRepPar}) that
\begin{align}
	YX^{-\top} &= e^{i\omega((0,1),(1,0))} YX =\zeta_1 XY        \\
	Y Y^{-\top}         & = e^{i\omega((0,1),(0,1))}\mbm{1} =\zeta_2 \mbm{1}.
\end{align}
Note that the results $H^2_{\beta^P}\left(\mb{Z}_2^P,\msf{U}_1\right) = \mb{Z}_2$ and $H^2\left(\mb{Z}_2\times\mb{Z}_2,\msf{U}_1\right) = \mb{Z}_2$ are well known, corresponding to the existence of one non-trivial SPT phase under either parity, or under an on-site $\mb{Z}_2\times \mb{Z}_2$ symmetry. The fact that an a single on-site $\mb{Z}_2$ symmetry (which in itself does exhibit non-trivial projective representations) combined with the parity $\mb{Z}_2^P$ leads to a richer structure of SPT phases, as expressed by $H^2_{\beta^P}\left(\mb{Z}_2\rtimes\mb{Z}_2^P,\msf{U}_1\right)=\mb{Z}_2^{\times 2}$ is interesting. Using the choice of cocycles in Eq.~
\eqref{eq:CohomologyZ2Z2P} and the construction from Section~\ref{sec:ansatz}, explicit examples can be constructed for these different phases.

\section{Time reversal \& lattice symmetries}
\label{sec:time}
\emph{In this section we study time-reversal symmetry and time-reversal combined with translations over one site~\cite{wigner2012group}. The most important feature of time-reversal is that it is an anti-unitary transformation and can hence be written as $T=UK$ where $U$ is a unitary and $K$ denotes complex conjugation in a certain basis. $T$ can be represented linearly or projectively, depending on whether $U\overline{U}=\pm\mbm{1}$.\\
	We revisit the work by Chen et al. on the linear implementation of time-reversal in MPS and identify the corresponding SPT classification~\cite{Chen2011a}. We demonstrate that injective MPS can not be invariant under the projective representation of $T$, a tensor network manifestation of the Lieb-Schultz-Mattis theorem~\cite{lieb1961}. Finally we prove that time-reversal combined with a shift over one lattice site does not give rise to non-trivial SPT order and construct a canonical form for the trivial phase.}

\Sep

\subsection{Time reversal in TI systems}
\label{subsec:TimeRev}
$\sim\mbf{T^2=\mbm{1}, U\overline{U}=\mbm{1}}\sim$ Consider the uniform ansatz (\ref{eq:UniformMPS}). Time-reversal symmetry can then be implemented as
\begin{equation}
	\sum_j U_{ij}\overline{A}^j = X^{-1}A^iX.
\end{equation}
Note that without loss of generality we don't need to consider a phase in this transformation because such a phase can be consistently absorbed in the MPS tensor $A^i$.  For an MPS tensor in left canonical form, $X$ can furthermore be chosen unitary. Doing a second time-reversal results in
\begin{equation}
	A^i = \overline{X}^{-1}X^{-1}A^iX\overline{X}.
\end{equation}
By virtue of the fundamental theorem we have that $X\overline{X}=e^{i\chi}\mbm{1}$, which, combined with unitarity of $X$, results in $e^{i\chi} = \pm 1$ and thus $X = \pm X^\top$.\par
If $X$ is symmetric, writing $X=VV^\top$ (where $V$ is unitary because $X$ is) allows us to bring the MPS in a canonical form by means of the gauge transformation $A^i \mapsto \tilde{A}^i = V^\dagger A^iV$, which now transforms according to
\begin{equation}
	\sum_j U_{ij}\overline{\tilde{A}}^j = \tilde{A}^i.
\end{equation}
For a skew-symmetric $X$ we write $X = V\Gamma V^\top$, $V$ again being unitary, from which it follows that $\tilde{A}^i = V^\dagger A^iV$ transforms according to a quaternionic representation under $T$, up to multiplication by $U_{ij}$~\footnote{Consider $A$ transforming in a quaternionic representation according to $\overline{A}=X^{-1}AX$, where $X$ is skew-symmetric and unitary. $X$ can be brought in a skew-symmetric tridiagonal form $W$ by an orthogonal matrix $Q$: $W=Q XQ^\top$. Unitarity of $W$ implies that $X$ can be written as $X=\tilde{Q}\Gamma \tilde{Q}^\top$ for an orthogonal $\tilde{Q}$, showing that $\tilde{Q}^\top A\tilde{Q}$ transforms as $\overline{\tilde{Q}^\top A\tilde{Q}}\mapsto\Gamma^{-1} \tilde{Q}^\top A\tilde{Q}\Gamma$.}:
\begin{equation}
	\sum_j U_{ij} \overline{\tilde{A}^j} = \Gamma^{-1} \tilde{A}^i \Gamma.
\end{equation}
The cohomological classification corresponds to $H^1_{\alpha^T}\left(\mb{Z}_2^T,\msf{U}_1\right)\times H^2_{\beta^T}\left(\mb{Z}_2^T,\msf{U}_1\right)=\left\{e\right\}\times \mb{Z}_2$, as follows from the Smith normal form (Section~\ref{sec:SmithNormal}). The $\mb{Z}_2$ is understood as $X$ being (skew-)symmetric.

We can now also show that the entanglement spectrum in case of the non-trivial SPT phase for which $X= -X^\top$ is at least doubly degenerate~\cite{Pollmann2010}. Consider therefore the unique leading right eigenvector $\rho$ of the transfer matrix $\mb{E}$. In that case $\rho$ interpreted as a $D\times D$ matrix is (Hermitian) positive semidefinite by virtue of the quantum Perron-Frobenius theorem~\cite{albeverio1978frobenius, wolf2012quantum}. Consider some eigenvector $\mbf{x}$ of $\rho$ with positive eigenvalue $\lambda$, then by virtue of $X\overline{\rho}X^\dagger = \rho$ (proven in Appendix~\ref{App:Proof}), ${\mbf{x}}^\top X^\dagger$ is a left eigenvector of $\rho$ with the same eigenvalue $\lambda$. However, using $X^\top=-X$ it follows that $\Tr\left(X^\dagger \left(\mbf{x}\otimes\mbf{x}\right) \right)=0$, or in other words that $\mbf{x}^\top X^\dagger$ and $\mbf{x}$ are orthogonal eigenvectors belonging to the same eigenvalue $\lambda$.

$\sim\mbf{T^2=-\mbm{1}, U\overline{U}=-\mbm{1}}\sim$ This case is relevant for eg. $U=\sigma_Y$, which comes into play in the implementation of time-reversal symmetry on spin 1/2 particles~\cite{wigner2012group}. Let us not restrict to this particular example and consider a general unitary $U$ satisfying the aforementioned property $U\overline{U}=-\mbm{1}$. Starting again from the uniform ansatz (\ref{eq:UniformMPS}), considering a projective implementation of time reversal and applying it twice leads to:
\begin{equation}
	A^i \overset{T}{\mapsto} \sum_{j}U_{ij}\overline{A}^j \overset{T}{\mapsto} -A^i \overset{!}{=} A^i,
\end{equation}
from which we conclude that translationally invariant injective MPS cannot transform projectively under time reversal symmetry. This can be understood as a tensor network interpretation of the celebrated Lieb-Schultz-Mattis theorem~\cite{lieb1961} that dictates that the ground state of a system of half-integer spins ––in which case time-reversal acts projectively–- should be either symmetry broken (in contradiction with the assumption that the MPS is symmetric under time-reversal) or gapless (in which case the matrix product ansatz does not provide a good description).

\subsection{Time reversal combined with a one site shift}
We can now break translation invariance by considering following ansatz and imposing time-reversal symmetry up to a shift over one site:
\begin{equation}
	\includegraphics[valign=c]{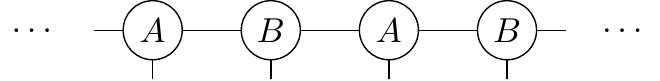}
	.
\end{equation}
The transformation of the tensors then reads
\begin{equation}
	\begin{cases}
		\sum_j U_{ij}\overline{A}^j & = X^{-1}B^iW  \\
		\sum_j U_{ij}\overline{B}^j & = W^{-1}A^iX
	\end{cases}
	,
\end{equation}
where $U\overline{U}=\mbm{1}$. Note that all phases can be absorbed in the tensors and gauge transformations.

A second transformation results in
\begin{equation}
	\begin{cases}
		A^i & = \overline{X}^{-1}W^{-1}A^iX\overline{W} \\
		B^i & = \overline{W}^{-1}X^{-1}B^iW\overline{X}.
	\end{cases}
\end{equation}
From blocking two tensors we conclude that $W\overline{X} = e^{i\chi}\mbm{1}$. Hence, we can absorb a factor $e^{i\chi/2}$ in both $\overline{X}$ and $W$ such that they become inverses. In conclusion, there is no non-trivial SPT phase and a canonical form is obtained by writing
\begin{equation}
	B^i = \sum_j U_{ij}X\overline{A}^j\overline{X}
\end{equation}
and defining $C^i=\overline{X}A^i$:
\begin{equation}
	\includegraphics[valign=c]{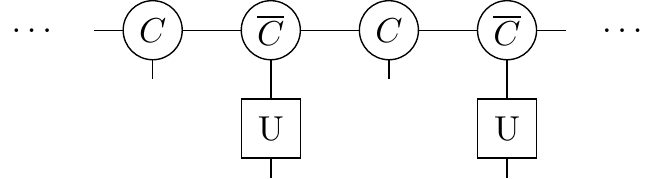}.
\end{equation}

\section{Conclusions and Outlook}
In this work we showed that quasi-one-dimensional spatial symmetries can protect non-trivial SPT phases in quantum spin chains represented by matrix product states. We identified each of these phases by invoking the symmetries on injective MPS and identifying the topologically distinct ways in which these symmetries can be represented by the local tensors. For most of these phases we constructed canonical MPS ans\"atze that are manifestly invariant under the considered symmetries. Finally, we revisited the SPT classification in case of time-reversal symmetry and showed that time-reversal combined with a translation over one lattice site does not give rise to non-trivial phases.

A natural extension of this work would be to consider the classification of two-dimensional SPT phases protected by space group symmetries. The two-dimensional space groups are known as the wallpaper groups, of which there are seventeen. In this case the relevant tensor network states are the projected entangled-pair states (PEPS), which form the natural two-dimensional generalization of the MPS considered here. We expect that similarly as in the one-dimensional case, imposing the spatial symmetries directly on the local tensors will also reveal topological obstructions. For each of these phases, canonical ans\"atze could be constructed that might prove very useful in numerical simulations of physical systems and materials in which these spatial symmetries are ubiquitous. The framework to investigate these spatial symmetries was laid out in~\cite{ThorgrenElse}, where it was called the \emph{crystalline equivalence principle}. This principle states that the classification of phases protected by a spatial symmetry group $\msf{G}$ is the same as that of the SPT phases with $\msf{G}$ as global on-site symmetry but acting in a `twisted' way, where orientation-reversing symmetry actions correspond to anti-unitary operators and thus to non-trivial group actions. In the tensor network framework this result was also obtained in~\cite{jiang2015,JiangRan}.

In particular, it would be interesting to investigate whether some of the symmetry transformations in 2D admit an implementation on the virtual level as string-like matrix product operators (MPOs). The physical application of the symmetry is then `gauged away' by pulling these MPOs through the lattice. Similarly, it might be interesting to demonstrate that also time-reversal, which, because of the complex conjugation, contains a priori a very non-local symmetry, can be implemented using an MPO of finite bond dimension. We plan to investigate this in future work.

\section*{Acknowledgments}
We would like to thank Robijn Vanhove for insightful comments regarding the extension to two dimensions, as well as Rui-Zhen Huang for many fruitful discussions on the topic of this work. B.V.-D.C. is supported by a Ph.D. fellowship from Bijzonder Onderzoeksfonds (BOF). This work has received funding from the European Research Council (ERC) under the European Unions Horizon 2020 research and innovation programme (grant agreements No 715861 (ERQUAF) and 647905 (QUTE)), and from Research Foundation Flanders (FWO) via grant GOE1520N.

\nocite{*}
%


\appendix
\section{Proof \texorpdfstring{of $X\overline{\rho}X^\dagger=\rho$}{}}
\label{App:Proof}
In this appendix we prove the identity $X\overline{\rho}X^\dagger=\rho$ which was used in section~\ref{subsec:TimeRev}. Hereto we start from the fact that $\rho$ was defined as the unique right eigenvector of the transfer matrix $\mb{E}$ (\ref{eq:TransferM}) corresponding to the eigenvalue one. Taking the complex conjugate of the eigenvalue equation and exploiting unitarity of $U$, we can show that $X^\dagger\overline{\rho}X$ is also a right eigenvector with eigenvalue one which because of injectivity and thus non-degeneracy of this eigenvalue has to be equal to $\rho$: $X\overline{\rho}X^\dagger=\rho$. Pictorially:

\begin{alignat}{2}
	&
	\quad
	\includegraphics[valign=c]{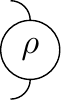}
	&   & =
	\includegraphics[valign=c]{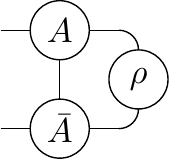}
	\\
	\iff &
	\quad
	\includegraphics[valign=c]{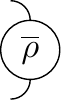}
	&   & =
	\quad
	\includegraphics[valign=c]{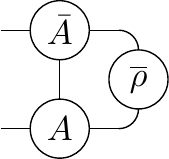}
	\\
	\iff &
	&   & =
	\includegraphics[valign=c]{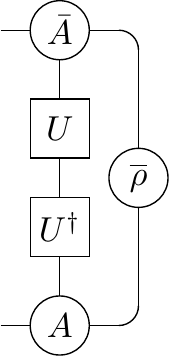}
	\\
	\iff &
	&   & =
	\includegraphics[valign=c]{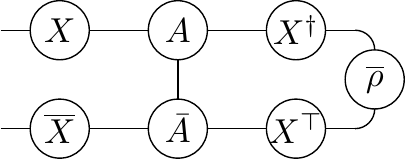}
	\\
	\iff &
	\quad
	\includegraphics[valign=c]{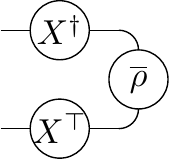}
	&   & =
	\includegraphics[valign=c]{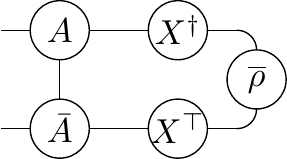}
\end{alignat}
\onecolumngrid

\section{Example: 2-cocycles of \texorpdfstring{$\mb{Z}_2^P$}{(ℤ₂)ᴾ}}\label{App:example}

In this appendix, we briefly illuminate the algorithm laid out in Sec.~\ref{sec:SmithNormal} for computing explicit cocycle representatives. We focus on one of the smallest interesting examples, $H^2_{\beta^P}\left(\mb{Z}_2^P,\msf{U}_1\right)$, which has a non-trivial group action. The cocycle equations to be solved thus read
\begin{equation}
	\omega(g,h) + \omega(gh,k) = \omega(g,hk) + \beta^P_g\left(\omega(h,k)\right) \mod 2\pi,
\end{equation}
where $\beta^P_g$ is multiplying with $-1$ when $g$ is the non-trivial element of $\mb{Z}_2^P$ and the identity otherwise.

The first step of the algorithm consists of filling up the $\Omega^{(2,\beta^P)}$-matrix (\ref{eq:System}), taking into account the non-trivial group action. Denoting the group elements of $\mb{Z}_2^P$ by $\{0,1\}$, we obtain:
\begin{align}
	\Omega^{(2,\beta^P)} = \bordermatrix{
		& (0;0) & (0;1) & (1;0) & (1;1) \cr
		(0;0;0) & 0     & 0     & 0     & 0     \cr
		(0;0;1) & -1    & 1     & 0     & 0     \cr
		(0;1;0) & 0     & 0     & 0     & 0     \cr
		(0;1;1) & 1     & -1    & 0     & 0     \cr
		(1;0;0) & -1    & 0     & -1    & 0     \cr
		(1;0;1) & 0     & -1    & -1    & 0     \cr
		(1;1;0) & -1    & 0     & -1    & 0     \cr
		(1;1;1) & 0     & -1    & 1     & -2    \cr
	}
\end{align}
This matrix is then written in Smith normal form as $P\Omega^{(2,\beta^P)} R=\Lambda$. The basis transformation $P$ contains only integers and can be discarded in solving the cocycle equations as these are solved modulo $2\pi$. The $R$ matrix reads

\begin{align}
	R & = \bordermatrix{
		& (0;0)            & (0;1) & (1;0) & (1;1) \cr
		& 1                & 0     & 1     & 0 \cr
		& 0                & 1     & 1     & 0 \cr
		& 0                & 0     & 1     & -1 \cr
		& 0                & 0     & 0     & 1 \cr
	}.
\end{align}
From the $\Lambda$ matrix,
\begin{align}
	\Lambda & = \bordermatrix{
		&                  &   &   & \cr
		& 1                & 0 & 0 & 0 \cr
		& 0                & 1 & 0 & 0 \cr
		& 0                & 0 & 2 & 0 \cr
		& 0                & 0 & 0 & 0 \cr
		& 0                & 0 & 0 & 0 \cr
		& 0                & 0 & 0 & 0 \cr
		& 0                & 0 & 0 & 0 \cr
		& 0                & 0 & 0 & 0 \cr
	},
\end{align}
the group cohomology is immediately found to be $H^2_{\beta^P}\left(\mb{Z}_2^P,\msf{U}_1\right)=\mb{Z}_2$: the $1$ diagonal entries corresponds to trivial cocycles, whereas the entry $2$ corresponds to a non-trivial cocycle. We then compute
\begin{align}
	R^{-1}\Lambda^+\vec{\nu} & = \bordermatrix{
		& \cr
		(0;0)                    & \nu _1-\frac{\nu _3}{2} \cr
		(0;1)                    & \nu _2-\frac{\nu _3}{2} \cr
		(1;0)                    & \frac{\nu _3}{2} \cr
		(1;1)                    & 0 \cr
	}\label{eq:CocycleZ2P}
\end{align}
where $\vec{\nu}$ is an arbitrary integer vector. The non-trivial cocycle valued in $\msf{U}_1$ that generates the $\mb{Z}_2$ cohomology group is then found from (\ref{eq:CocycleZ2P}) by choosing $\nu_3=1$ and reads
\begin{align}
	\vec{\omega} & = \bordermatrix{
		& \cr
		(0;0)        & -1	\cr
		(0;1)        & -1	\cr
		(1;0)        & -1 	\cr
		(1;1)        & 1 	\cr
	}.
\end{align}

\end{document}